\documentclass[fleqn,twoside]{article}
\usepackage[headings]{espcrc2}
\def\1{\mbox{l\hspace{-0.53em}1}}
\readRCS
$Id: espcrc2.tex,v 1.2 2004/02/24 11:22:11 spepping Exp $
\usepackage{graphicx}
\usepackage[figuresright]{rotating}

\newcommand{\AmS}{{\protect\the\textfont2
  A\kern-.1667em\lower.5ex\hbox{M}\kern-.125emS}}

\title{Excited $[{\bf 70},\ell^+]$ baryon resonances in large $N_c$ QCD}

\author{N. Matagne\address[UNIF]{University of Li\`ege, Physics Department,\\
Institute of Physics, B.5, \\
Sart Tilman, B-4000 Li\`ege 1, Belgium\\}\thanks{E-mail address: nmatagne@ulg.ac.be} and Fl. Stancu\addressmark[UNIF]\thanks{E-mail address: fstancu@ulg.ac.be}}

\begin{document}

\begin{abstract}
We summarize results obtained in the $1/N_c$ expansion method
for the masses of baryon resonances belonging to the $[{\bf 70},\ell^+]$
multiplet. They represent an extension of our previous studies from
two to three flavors. A better approach to mixed symmetric states
of any angular momentum and parity is 
also outlined.
\vspace{1pc}
\end{abstract}

\maketitle

\section{Introduction}
QCD is not a perturbative theory at low energy regime. Consequently, one does not know how to 
systematically study the structure of the ground  and excited baryon states except, in principle, for lattice calculations. 
Large $N_c$ QCD \cite{HOOFT,WITTEN} introduces a new perturbative series where the parameter is 
the inverse of the number of colors $N_c$.
The resulting $1/N_c$ expansion approach has proven to be an interesting way to study 
baryon spectroscopy.  

The success is due to the discovery in 1984, by Gervais and Sakita \cite{gervais} and independently, 
during the ninetieths, by 
Dashen and Manohar \cite{DM93}, from the study of baryon-meson scattering process in 
the $N_c \to \infty$ limit, that ground state baryons satisfy a contracted SU($2N_f)_c$ 
spin-flavor algebra where $N_f$ is the number of flavors. This means that 
when $N_c \to \infty $, ground state baryons form an infinite tower of degenerate states.
At $N_c \to \infty $, the SU($2N_f)$ algebra used in the constituent quark model becomes 
the SU($2N_f)_c$ algebra. One can therefore use SU($2N_f$) to classify large $N_c$
baryons and calculate matrix elements of various static observables. 
If SU($2N_f$) is broken the degenerate baryon states split at order $1/N_c$.

The first studies of excited baryons in the large $N_c$ limit suggested that constituent quark baryons 
can be reduced to a symmetric core composed of  $N_c-1$ ground state quarks and one 
excited quark. The advantage is that one can treat the core in the same way as a ground 
state baryon. However, with this approach, the SU($2N_f$) symmetry is broken at 
order $\mathcal{O}(N^0_c)$ instead of $\mathcal{O}(1/N_c)$ 
as for the ground state  \ \cite{Goi97}. This generates the conceptual problem 
that in the $N_c \to \infty$ limit, excited states do not form anymore an infinite tower 
of degenerate states. Furthermore, excited states are resonances and have widths of order 
$N_c^0$ \cite{cohen1}. Nevertheless,  baryons belonging to various SU(6) excited multiplets 
have been studied with success during the last ten years 
\cite{CGKM,PY,CCGL,CaCa98,BCCG,SGS,SCHAT,Pirjol:2003ye,GSS03,MS2,MS1,MS3}. 
More or less, all these studies consider excited baryons as bound states and 
split the SU($2N_f$) generators into two parts, one acting on the symmetric core and the
other on the excited quark. Accordingly, the number of invariant operators needed
in the description of observables becomes exceedingly large and the splitting starts 
at order $\mathcal{O}(N^0_c)$. Fortunately, these studies show that the $N^0_c$ breaking is small.

Recently, a new approach of the $[{\bf 70}, 1^-]$ multiplet solved this conceptual 
problem by removing  the splitting of generators and using 
orbital-flavor-spin wave functions respecting the permutation symmetry \cite{ms5}.
Still, the  excited baryons 
as considered as bound states.    

Below we shall analyze the $[{\bf 70},\ell^+]$ multiplets with $\ell$ = 0 or 2 
in the standard approach with splitting.

\section{The mass operator}
The constituent quark model suggests that $[{\bf 70},\ell^+]$ baryons are composed of one or 
two excited quarks and $\mathcal{O}(N_c)$ ground state quarks. The standard procedure, used in 
Ref. \cite{MS3}, for calculating the mass spectrum is to
reduce the wave function to that of a product of a symmetric orbital and a symmetric flavor-spin wave function
 for the core
of $N_c-1$ quarks times the wave function of one excited quark.
This implies that the total orbital-flavor-spin wave function is truncated to
a single term, described by the product of two Young tableaux, each with the excited quark in the 
second row.  
Many others terms, related to Young tableaux with the excited quark in the first row, are neglected \cite{ms5}.
At the same time each SU(6) and SO(3) generator is splitted into two terms, one acting on the
core and the other on the excited quark.  We assume that baryons are bound states. 

The mass operator must be rotationally invariant, parity and time reversal even. For the $[{\bf 70}, \ell^+]$ multiplet it has the following structure:
\begin{equation}
 M_{[{\bf 70},\ell^+]}=\sum^6_{i=1}c_iO_i +d_1B_1+d_2B_2+d_4B_4,
\end{equation}
where $O_i$  are rotational invariants and SU(3)-flavor scalars, the operators $B_i$ provide SU(3) 
breaking and  are defined to have non-vanishing matrix elements for strange baryons only and the 
coefficients $c_i$ and $d_i$, which encode the QCD dynamics, are unknown. One can obtain these coefficients from a fit to experimental data. As the number of data is limited, we have to make a selection among all the possible operators.

The first column of Table \ref{operators} shows the list of operators chosen for this study
as thought to be the most dominant one. 
The choice is based on previous studies \cite{CCGL,SGS,MS1}. They are composed of 
SU(6)$\times$SO(3) generators. Generators $S^i_c,\ T^a_c,\ G^{ia}_c$ refer to the core part 
of the wave function and $s^{i},\ t^a,\ g^{ia}$ and $\ell_q^{i}$ act on the excited quark. Besides the SO(3) generators which are of rank $k=1$ we also need to introduce the rank $k=2$ tensor operator 
  \begin{equation}
   \ell^{(2)ij}_q=\frac{1}{2}\left\{\ell^i,\ell^j\right\}-\frac{1}{3}\delta_{i,-j}\vec{\ell}\cdot \vec{\ell}.
  \end{equation}
  
  For deriving the matrix elements of the operators $O_i$ and $B_i$,  we first needed to calculate 
  the matrix elements of the core generators, the problem being trivial
   for the excited quark  (single particle operators).  It is not always an obvious task. 
   To obtain the matrix elements, we have used a generalized Wigner-Eckart theorem and then 
   derived SU(6) isoscalar factors. We have obtained analytic expressions for these isoscalar 
   factors for SU(6) symmetric wave functions $|[N_c] \rangle $  containing any number
   of quarks \cite{MS4}. The mixed symmetric case needs considerable 
   group theory work. 
   To our knowledge, isoscalar factors 
   for mixed symmetric states $|[N_c - 1, 1] \rangle $ which can be applied to baryons composed 
   of  $N_c$ quarks \cite{ms5} for $N_f=3$ are yet inexistent. That is why, presently, 
   it is not possible to treat the $[{\bf 70},\ell^+]$ multiplet composed of strange baryons 
   without simplifying the baryon wave function and splitting the SU(6) generators. 
\begin{table*}[htb]
\begin{center}
\caption{List of operators and the coefficients resulting from the fit with 
$\chi^2_{\rm dof}  \simeq 1.0$ for the $[{\bf 70},\ell^+]$ multiplets ($\ell = 0$ and 2).}
{\label{operators}
\renewcommand{\arraystretch}{1.5} 
\begin{tabular}{llrrl}
\hline
Operator & \multicolumn{4}{c}{Fitted coef. (MeV)}\\
\hline
$O_1 = N_c \ \1$                                    & \ \ \ $c_1 =  $  & 556 & $\pm$ & 11       \\
$O_2 = \ell_q^i s^i$                                & \ \ \ $c_2 =  $  & -43 & $\pm$ & 47    \\
$O_3 = \frac{3}{N_c}\ell^{(2)ij}_{q}g^{ia}G_c^{ja}$ & \ \ \ $c_3 =  $  & -85 & $\pm$ & 72  \\
$O_4 = \frac{4}{N_c+1} \ell^i t^a G_c^{ia}$         & \ \ \            &     &       &     \\
$O_5 = \frac{1}{N_c}(S_c^iS_c^i+s^iS_c^i)$          & \ \ \ $c_5 =  $  & 253 & $\pm$ & 57  \\
$O_6 = \frac{1}{N_c}t^aT_c^a$                       & \ \ \ $c_6 =  $  & -25 & $\pm$ & 86    \\ 
\hline
$B_1 = t^8-\frac{1}{2\sqrt{3}N_c}O_1$               & \ \ \ $d_1 =  $  & 365 & $\pm$ & 169 \\
$B_2 = T_c^8-\frac{N_c-1}{2\sqrt{3}N_c}O_1$         & \ \ \ $d_2 =  $  &-293 & $\pm$ & 54 \\
$B_4 = 3 \ell^i_q g^{i8}- \frac{\sqrt{3}}{2}O_2$    & \ \ \            &     &       & \vspace{0.2cm}\\
\hline
\end{tabular}}
\end{center}
\end{table*}

In Table \ref{operators}, $O_1$ is the SU(6) scalar operator linear in $N_c$. $O_2$ and $O_5$ 
are the dominant part of the spin-orbit and spin-spin operators respectively. The first, which 
acts only on the excited quark, is of order $N_c^0$ but the two-body spin-spin operator is of 
order $N_c^{-1}$. The operators $O_3$ and $O_4$ are of order $N_c^0$ due to the presence of the 
SU(6) generator $G_c^{ia}$ which sums coherently. $O_6$ represents the isospin-isospin operator,
 having matrix elements of order $N_c^0$ due to  $T_c^a$ which sums coherently  too.
 
 As already mentioned, the operators $B_i$ break the SU(3)-flavor symmetry. The operators 
 $B_1$ and $B_2$ are the standard breaking operators while $B_4$ is directly related to the 
 spin-orbit splitting. They break the SU(3)-flavor symmetry to first order.

\section{Results}
The second column of Table \ref{operators} gives the values of the coefficients $c_i$ and 
$d_i$ resulting form the fit. The $\chi^2_{\mathrm{dof}}$ obtained is 1.0. Details concerning 
the calculations and data used in  the fit are presented elsewhere \cite{MS3}. One can see that 
the first order operator $O_1$ and the spin-spin operator $O_5$ are the most dominant 
ones, \emph{i.e.} $c_1$ and $c_5$ are large. The spin-orbit coefficient is negative, 
at variance with  previous studies \cite{MS2} but remains small in absolute value. The 
coefficient $c_3$ is twice smaller in absolute value as compared to that of Ref. \cite{MS2}. 
We had to exclude the operator $O_4$ from the fit because it considerably deteriorated  
the $\chi^2_{\mathrm{dof}}$. This study suggests that the isospin-isospin operator does not 
play an important role, the coefficient $c_6$ being small. Possibly, this is a consequence of
the splitting of generators and of the truncation of the baryon wave function. 
In the improved approach of Ref. \cite{ms5}, applied to  
 the analysis of the $[{\bf 70},1^-]$ multiplet,
the isospin term appears to have the same importance as the spin term. 

The SU(3)-flavor breaking operators play an important dynamical role as it can be seen from 
the values of the  coefficients $d_1$ and $d_2$. As all the matrix elements of $B_4$ cancel out 
for the available resonances,  it was not possible for us to obtain an estimation of $d_4$. 

\begin{figure}[h!]
\centerline{\includegraphics[width=7.5cm,keepaspectratio]{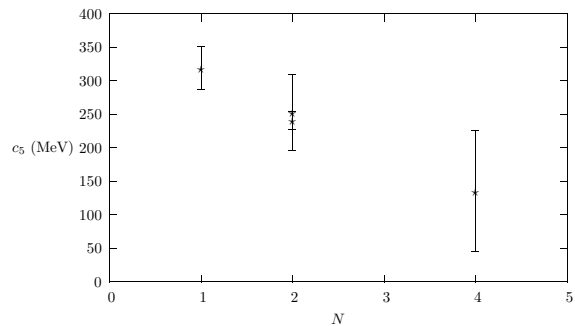}}
\caption{Evolution of the coefficient $c_5$ with the excitation energy corresponding to $N=1, 2$ and 4 bands in a harmonic oscillator notation.}
\label{c2}
\end{figure}

Figure  \ref{c2} shows the evolution of the spin-spin dynamical coefficient $c_5$ with the 
excitation energy.  Here we have collected the presently known values with error bars for the 
orbitally excited states studied so far in the large $N_c$ expansion: 
$N=1$, Ref. \cite{SGS}, $N=2$ (lower value \cite{GSS03}, upper value \cite{MS3}) 
and $N=4$, Ref. \cite{MS1}. This figure suggests that at large excitations, the spin-spin 
contribution vanishes.

\section{Conclusions}
We have shown results obtained for the masses of resonances which we identified as belonging to 
the $[{\bf 70},0^+]$ and $[{\bf 70},2^+]$ multiplets. The present results confirm the dependence
 of the coefficients $c_1$, $c_2$ and $c_5$ as a function of excitation energy, namely that the 
 contributions of the spin-dependent terms decrease with energy and eventually vanish at very 
 large excitations \ \cite{MS2}. The analysis of $[{\bf 70},\ell^+]$ remains open.
 More and better experimental data are needed to clarify the role of various terms contributing 
 to the mass operator of the $[{\bf 70},\ell^+]$ multiplet. A new study should be made without 
 truncating the baryon wave function.

\section*{Acknowledgments}
The work of one of us (N. M.) was supported by the Institut Interuniversitaire des Sciences Nucl\'eraires (Belgium).

\end{document}